\documentclass[Afour,Sageh,times]{sagej}

\usepackage{moreverb,url}
\usepackage{tcolorbox}
\usepackage[colorlinks,bookmarksopen,bookmarksnumbered,citecolor=red,urlcolor=red]{hyperref}
\usepackage{siunitx}
\usepackage{multirow}

\newcommand\BibTeX{{\rmfamily B\kern-.05em \textsc{i\kern-.025em b}\kern-.08em
T\kern-.1667em\lower.7ex\hbox{E}\kern-.125emX}}

\newcommand{\vect}[1]{\boldsymbol{#1}}

\def\volumeyear{2020}

\begin{document}

\runninghead{de Souza and Berger}

\title{Fallopian tube anatomy predicts pregnancy and pregnancy outcomes after tubal reversal surgery}

\author{Rafael S. de Souza\affilnum{1} and Gary S. Berger\affilnum{2}}

\affiliation{\affilnum{1}Key Laboratory for Research in Galaxies and Cosmology, Shanghai Astronomical Observatory, Chinese Academy of Sciences, 80 Nandan Road, Shanghai 200030, China\\
\affilnum{2}Department of Obstetrics and Gynaecology, University of North Carolina at Chapel Hill,  NC 27599-3255, USA}

\corrauth{Rafael S. de Souza, Gary S. Berger}
\email{drsouza@shao.ac.cn, drberger@chfc.net}

\begin{abstract}
We conducted this study to determine whether fallopian tube anatomy can predict the likelihood of pregnancy and pregnancy outcomes after tubal sterilization reversal. 
We built a flexible, non-parametric, multivariate model via generalized additive models to assess the effects of the following tubal parameters observed during tubal reparative surgery: tubal lengths; differences in tubal segment location and diameters at the anastomosis sites; and, fibrosis of the tubal muscularis.
In this study population, age and tubal length - in that order - were the primary factors predicting the likelihood of pregnancy. For pregnancy outcomes, tubal length was the most influential predictor of birth and ectopic pregnancy, while age was the primary predictor of miscarriage. Segment location and diameters contributed slightly to the odds of miscarriage and ectopic pregnancy. Tubal muscularis fibrosis had little apparent effect.
This study is the first to show that a statistical learning predictive model based on fallopian tube anatomy can predict pregnancy and pregnancy outcome probabilities after tubal reversal surgery.
\end{abstract}


\maketitle

\section{Introduction}
Female sterilization is the most common contraceptive method worldwide \cite{DANIELS2018}. Approximately 9.5 million US women rely on female sterilization for birth control  \cite{DANIELS2015}.  Although sterilization is intended to be permanent, postmodern society has experienced paradigmatic behavioral changes with increased rates of divorce and remarriage \cite{Randall1999}. In this context, many sterilized women have expressed regret and wish to have their fertility restored. The frequency rate of the request for tubal sterilization reversal  can be as high as 14.3\% \cite{CURTIS20062}.

Tubal reparative surgery is an effective treatment for many women desiring fertility restoration after a tubal ligation. Preoperative counseling, based on age and method of sterilization, can estimate the likelihood of pregnancy and pregnancy outcomes following tubal anastomosis \cite{BERGER2016}.  After tubal reparative surgery, patients usually inquire about the condition of their fallopian tubes and how that affects their prognosis. We undertook this study to answer this question.

In this report, we present a model for predicting pregnancy and pregnancy outcome probabilities based on a woman's age and tubal anatomy as observed during tubal reversal surgery.

\section{Methods}

This section introduces the dataset utilized
in this study; the pre-processing of our data; and the 
statistical models employed for the analysis of pregnancy and pregnancy outcomes.

\subsection{Study design and population}
\label{sec:data}

This report makes  use of a dataset from a prospective observational study of women who had tubal reconstructive surgeries. The operations were performed in an outpatient surgical center in Chapel Hill, NC, USA, from January 2000 to June 2013. The surgical and study methods have been described in detail previously  \cite{BERGER2016}.  The University of North Carolina at Chapel Hill IRB and Office of Human Research Ethics gave this study exempt status (IRB Number 14-1783) as a quality improvement study, meaning that written consent was not required.

Tubal anatomy was assessed at surgery for both segments joined at the anastomosis site for right and left fallopian tubes independently. The measurements included: length; location (interstitial, isthmic, ampullary, infundibular, or fimbrial); diameter; and degree of fibrosis. Location differences were analyzed numerically by segment position difference (SPD). For example, if the segments were identical such as isthmic-isthmic or ampullary-ampullary, the location difference was 0; isthmic-ampullary or ampullary-infundibular difference was 1, and so on, Diameters of the anastomosed segments were categorized as similar, somewhat dissimilar, or dissimilar. Fibrosis of the tubal muscularis, assessed visually and by palpation, was recorded as none, mild, moderate, or severe.

Among all 9,669 women in the tubal surgery database, we retrieved information from the those who had bilateral tubotubal anastomosis with complete information about all fallopian tube anatomic parameters and at least one year of follow-up after surgery.  These 5,682 women comprised the study population for the present analysis.

\begin{figure} 
\includegraphics[width=0.95\linewidth]{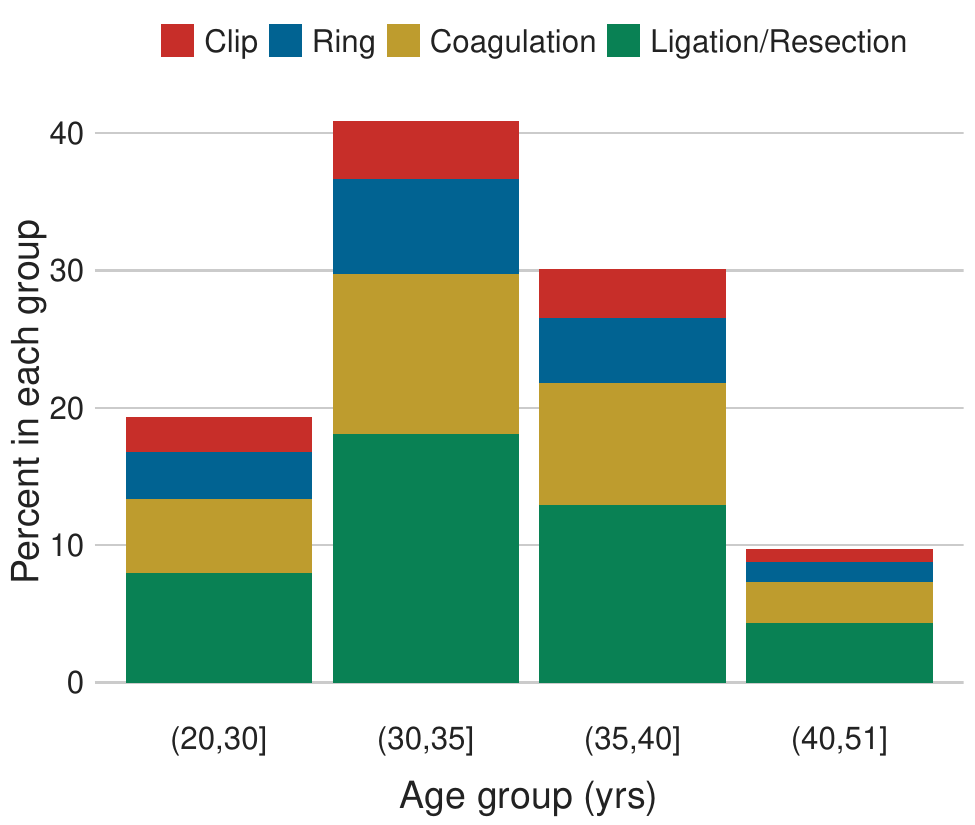}
\caption{Age group distribution of 5682 women in which  19.3\% were younger than 30 yrs, 40.9\%  between 30-35 yrs, 30.1\% between 35-40 yrs, and 9.7\% older than 40 yrs. Sterilization methods are color-coded, with ligation/resection representing the most common method across all ages.}
\label{fig:age}
\end{figure}

The women's ages at the time of reconstructive surgery ranged from 20 to 51 years.  The  distribution of age groups and sterilization methods within each age group are shown in Fig. \ref{fig:age}.  In this study population,  19.3\% were younger than 30 yrs, 40.9\% were 30-35 yrs, 30.1\% were 35-40 yrs, and 9.7\% were 40 yrs of age or older. Across all age groups, the most common method of sterilization was ligation/resection, followed in order by coagulation and by mechanical methods (rings or clips).  

\subsection{Feature selection}
\label{sec:fs}

For every record, we extracted the woman's age 
in combination with the following anatomic properties for both left and right fallopian tubes: tubal length after anastomosis, specific tubal segments rejoined, diameters of the two segments at the anastomosis site, and fibrosis of the tubal muscularis.

Which one of a woman's fallopian tubes results in a given pregnancy is unknown. Therefore, we emulated the randomness of the choice between left and right sides by attributing to each woman either left or right tubal anatomic properties following a Bernoulli process with 50\% probability to chose one side or another. 

\begin{figure} 
\includegraphics[width=0.975\linewidth]{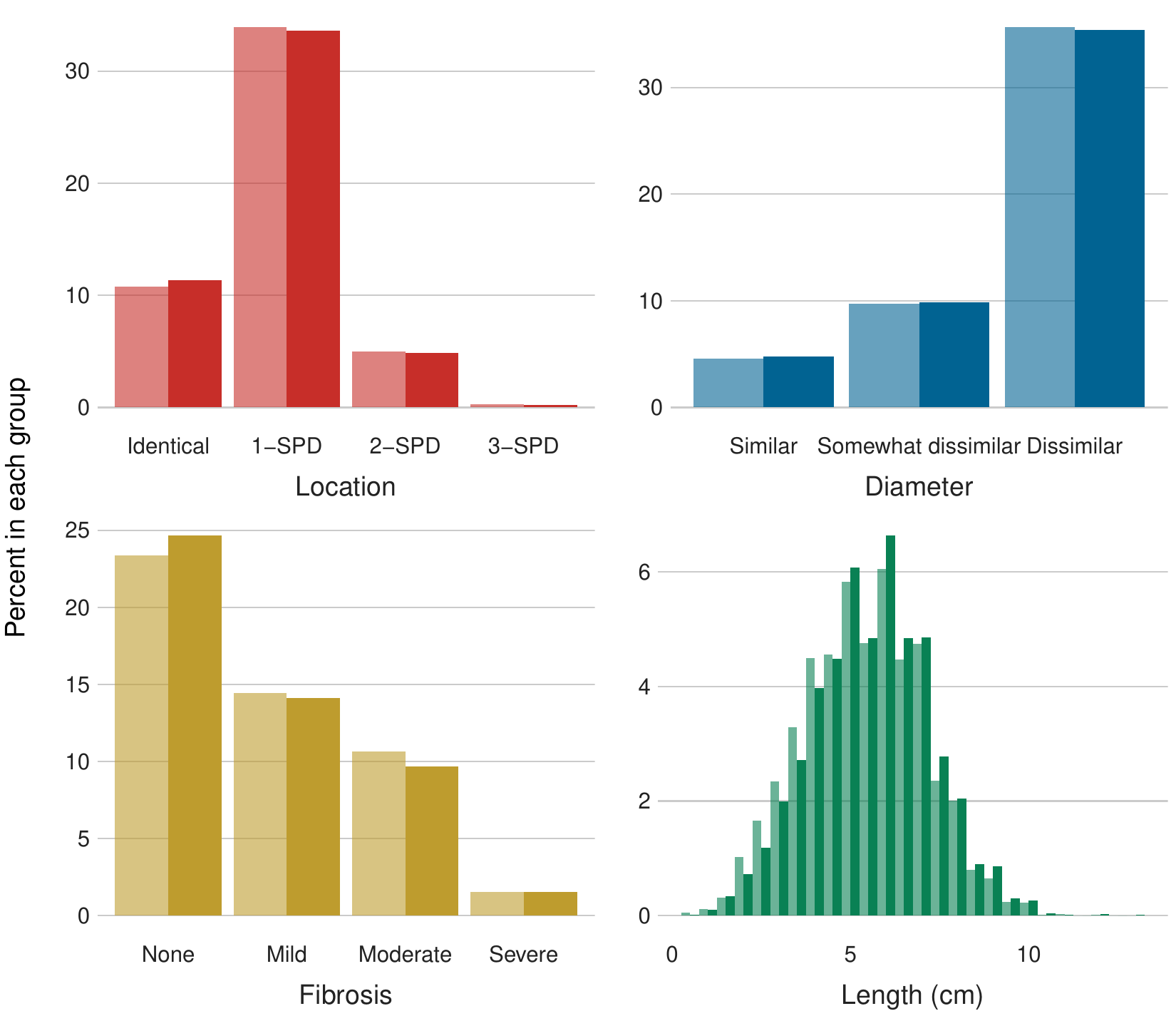}
\caption{Percent in each group for anatomic properties: Location differences, upper left panel; Diameter differences, upper right panel; Fibrosis, lower left panel; Tubal length, lower right panel. In each panel the lighter bars represent the left tube, and the darker bars the right tube. For Location, SPD stands for segment position difference.}
\label{fig:ana}
\end{figure}

A Bernoulli distribution describes a sequence of independent experiments (trials) each of which has only two possible outcomes $\{0,1\} $. In the specific case of interest here, one can think of the choice between left and right sides as binary data which is either left = 0,  or right = 1. Fig. \ref{fig:ana} displays the distribution of tubal anatomic properties for the study population in terms of left and right features. Simple visual inspection does not reveal any directional bias towards a particular side.

Every record was then represented as a 5-dimensional feature vector with 2 numeric properties (age, tubal length) and 3 categorical properties (location differences, diameter differences, extent of fibrosis). For the response variable, there are two parts of the analysis. The first concerns pregnancy prevalence and retrieves binary information: the woman became pregnant or not. The second part exploits each pregnancy outcome: birth, miscarriage, ectopic, or ongoing at time of last contact.

\begin{figure} 
\includegraphics[width=0.95\linewidth]{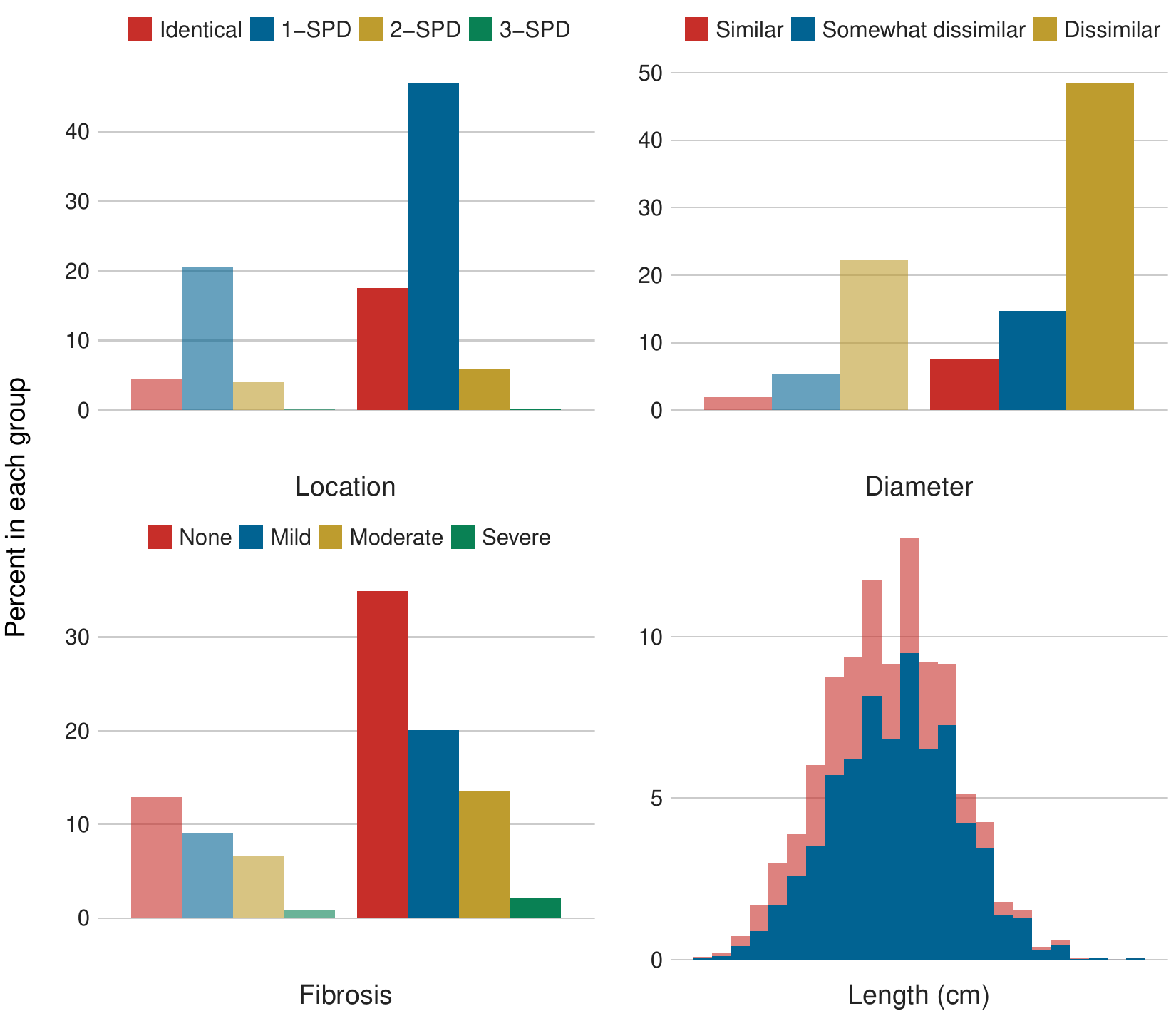}
\caption{Percent in each group for anatomic properties: Location differences, upper left panel; Diameter differences, upper right panel; Fibrosis, lower left panel; Tubal length, lower right panel. In each panel the lighter bars represent the non-pregnant women, and the darker bars the pregnant ones. }
\label{fig:bar_preg}
\end{figure}

Fig. \ref{fig:bar_preg} shows the distribution of anatomic properties in terms of pregnancy occurrence. The median tubal length is slightly larger for women who became pregnant (5.5 cm) than those who did not (5 cm). Likewise,  there appears to be a preference towards anastomosis of segments of identical location or 1-segment position difference and for none or mild fibrosis of the tubal muscularis among the pregnant women. Diameter differences do not show any visually detectable bias between pregnant and non-pregnant women. 

\begin{figure*} 
\includegraphics[width=0.325\linewidth]{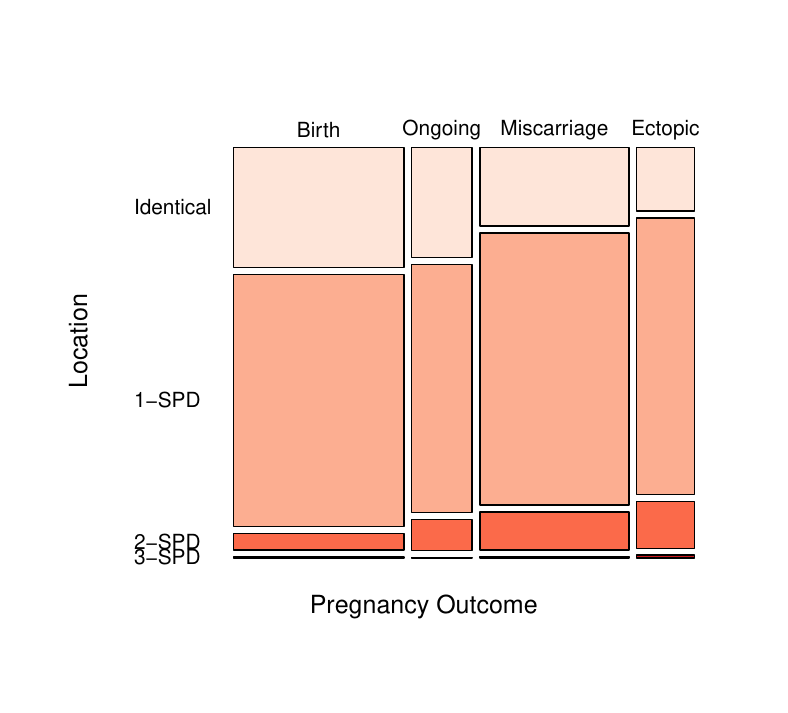}
\includegraphics[width=0.325\linewidth]{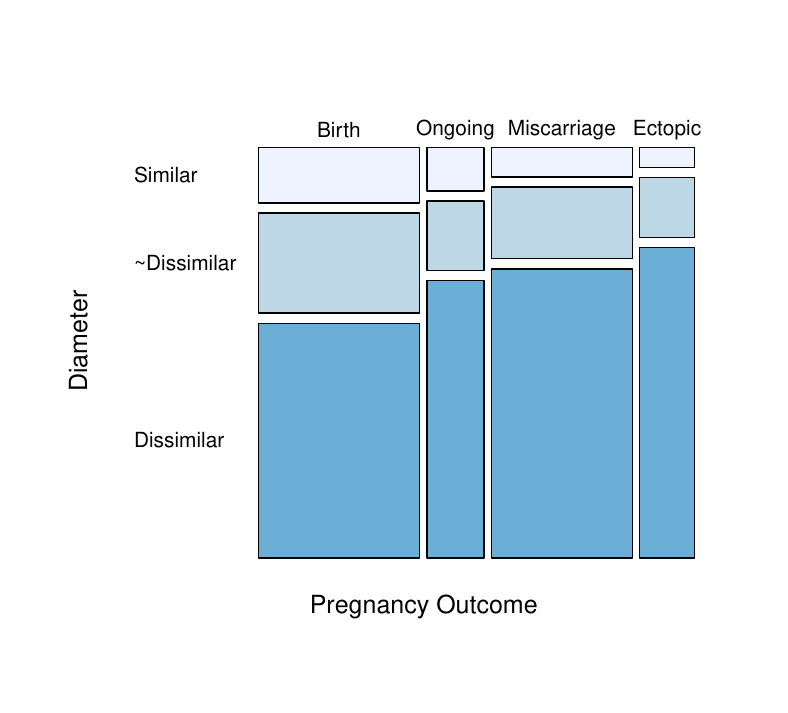}
\includegraphics[width=0.325\linewidth]{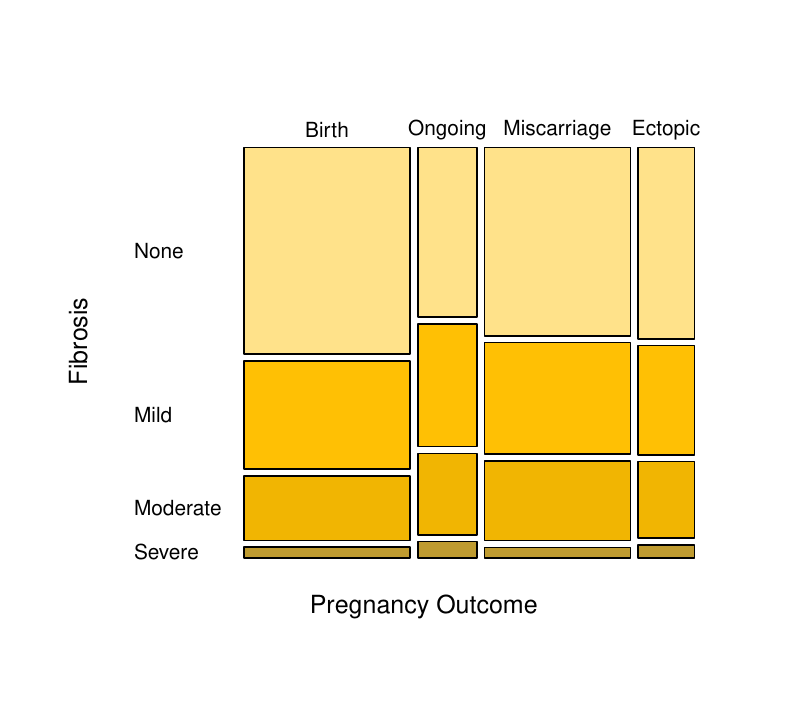}
\caption{Populations of different pregnancy outcomes in terms of categorical tubal anatomic properties. From left to right: Location, Diameter and Fibrosis properties. In each panel the width of the bar is proportional to the total size of the population.}
\label{fig:ana_mosaic}
\end{figure*}

Fig. \ref{fig:ana_mosaic} displays the distribution of anatomic properties in terms of pregnancy outcome. In each panel the width of the bar is proportional to the total size of the class population. Visual inspection suggests that the greater the differences in segment location and diameters, the greater the likelihood of miscarriage and ectopic pregnancy. The presence or degree of fibrosis has little apparent effect on pregnancy outcome.

In what follows, we introduce a more quantitative analysis of all properties simultaneously for their influence on pregnancy and outcome likelihoods.

\subsection{Generalized Additive Models}
\label{sec:CEL}

The simple \textit{linear regression} model, although ubiquitous, falls short when the data to be modeled come from \textit{exponential family} distributions other than the Normal/Gaussian \cite{deSouza2015,deSouza2015b,Elliott2015,Hilbe2017}. 
For such  problems, there is a solution  known as generalized linear models (GLMs). Regression models in the class of GLMs \cite{nel72}, take a more general form than in ordinary linear regression:
\begin{align}
\label{eq:glm}
&Y_i \sim f(\mu_i, a(\phi) V(\mu_i)) , \notag \\
&g(\mu_i) = \eta_i,  \\ 
&\eta_i \equiv \boldsymbol{x}_i^T \boldsymbol{\beta} = \beta_0+\beta_1x_1+\cdots+\beta_kx_k. \notag 
\end{align}
In equation~(\ref{eq:glm}), $f$ denotes a response variable distribution from the exponential family (EF), $\mu_i$ is the response variable mean, $\phi$ is the EF dispersion parameter in the dispersion function $a(\cdot)$, $V(\mu_i)$ is the response variable variance function, $\eta_i$ is the linear predictor,  
$\boldsymbol{x}_i^T$
is a vector of explanatory variables (covariates or predictors), 
$\boldsymbol{\beta}$
is a vector of covariate coefficients, and $g(\cdot)$ is the link function, which connects $\mu_i$ to $\eta_i$.  If the response is Gaussian, then $g(\mu) = \mu$, and it recover the Gaussian, $\mathcal{N}$,   linear regression as a subset:
\begin{align}
\label{eq:lm}
&Y_i \sim \mathcal{N}(\mu_i, \sigma^2) , \notag \\
 &\mu_i =  \boldsymbol{x}_i^T \boldsymbol{\beta}.  
\end{align}

The methodology discussed herein focuses on two particular classes of GLMs known as logistic and multinomial regression, which are suitable for handling Bernoulli (or binomial) and multinomial/categorical data. 
The Bernoulli distribution is a particular case of the more general binomial
distribution, $\mathrm{Binomial}(n,p) =  \binom {n} {y} p^y (1-p)^{n-y}$, for
which $y$ is the number of successes ($y = 1$),   $n$ is the number of trials, and $p$ is the probability of success.  For the Bernoulli
distribution, $\mathrm{Bernoulli}(p) =  p^y (1-p)^{1-y}$, the number of trials, $n$, is set to 1. 

The natural link function for the Bernoulli distribution is
known as the logit link, which defines the logistic model:
\begin{equation}
\label{eq:logit}
g(\mu_i=p_i)= \mathrm{logit}(p_i) \equiv \log \left(\frac{p_i}{1-p_i}\right), 
\end{equation}
and ensures a bijection between
the $(-\infty,\infty)$ range of $\eta_i$, and the (0,1)
range of non-trivial probabilities for the
Bernoulli $p_i$.
The logistic model takes the form:
\begin{align}
&Y_i \sim \rm {Bern}(p_i), \notag \\
&\mathrm{logit}(p_i) =  \boldsymbol{x}_i^T \boldsymbol{\beta}.
\end{align}
A natural extension, in which more than two outcome values are possible,  is called multinomial (or categorical) logistic model: 

\begin{align}
&Y_i \sim \rm {Cat}(p_{i;k}), \notag \\
&\mathrm{logit}(p_{i;k}) = \boldsymbol{x}_i^T \boldsymbol{\beta_k},
\end{align}
where the categorical distribution corresponds to the multinomial where just one choice is made, in a similar matter as the Bernoulli is to the binomial.

For many instances, the functional formula in equation \ref{eq:glm} can
be quite restrictive. It may not account properly for non-linear functional relationships and lacks the flexibility
to detect local patterns in the data.
As for our analysis, we employed  an  important extension of the  GLM methodology  known as generalized additive models (GAM) \cite{Hast1990}.  The model, in this context, extends the linear limitation by assuming  the existence of an  unknown  functional relationship between $g(\mu)$ and a set of covariates $\vect x$.  In other words, $g(\mu) = f(\vect x) $ for unknown $f$.  GAMs assume that $f(\vect x ) = f_1( x_1)  + f_2(x_2) + \hdots + f_k( x_k)$, and  $f$ can be approximated by a given basis function, for which we employed a cubic spline for the age and tubal length covariates.
Throughout this work, we evaluate the GAM model using the implementations \cite{Thomas2010,Simon2016} within the {\sc R} language \cite{R}.

\subsection{Pregnancy Likelihood}

To model the pregnancy likelihood (PL), we adopted the following logistic model:
\begin{align}
\label{eq:model}
   &\rm PL_{i}\sim \rm {Bern}\left(p_{i}\right), \notag   \\
  &\mathrm{logit}(p_i) =  f(Age_i) + f(Length_i) + \\
  &  Location_i + Fibrosis_i + Diameter_i.\notag 
    \end{align}
 The  model reads as follows.  Each of the $i$-th  women in the dataset has its probability to get pregnant  modeled as a Bernoulli process, whose probability of success relates to the age and fallopian tube anatomy through a logit link function, $\eta_{i}$. The model assumes a smooth function of the age and tubal length, and accounts for the categorical variables location, fibrosis  and diameter.

\subsection{Outcome Likelihood}

The outcome likelihood (OL) is assumed to follow a categorical  model in the form:
\begin{align}
\label{eq:model2}
 &\rm OL_{i}\sim \rm {Cat}\left(p_{i;k}\right), \notag   \\
   &\mathrm{logit}(p_{i;k}) =  f(Age_i) + f(Length_i) + \\
  &  Location_i + Fibrosis_i + Diameter_i,\notag 
    \end{align}
where the  model reads as follows.  Each of the $i$-th  women in the dataset has its outcome likelihood probability, $p_{i;k}$, encoded by $k$ = 1$\dots$4 categories: birth, ongoing, miscarriage, and ectopic modeled as a multinomial process, in which the weighted probability of each outcome  relates to the age and fallopian tube anatomy through a logit link function. The model assumes a smooth function of the age and tubal length, and accounts for the categorical variables location, fibrosis and diameter.

\subsection{Cross-Entropy Loss}

The cross-entropy loss, $H_p$,    measures the performance of a classification model whose output is a probability value between 0 and 1 \cite{commenges2009}. The $H_p$ increases as the predicted probability diverges from the actual label. Hence, the lower the $H_p$, the better is the model's predictive capabilities. 

In the case of multi-classification problems,  with $\kappa$ classes, we estimate a loss for each class label per observation and sum the result:

\begin{equation}
H_{p} = - \frac{1}{N} \sum_{i=1}^{N}\sum_{j = 1}^{\kappa} y_{i,j}\log{p_{i,j}}, 
\end{equation}
where $\kappa$ is the number of classes (birth, miscarriage, ectopic, outgoing), y is a binary indicator if class label $j$ is the correct classification for a given observation $i$, $p$ is the predicted probability for a given class $j$. 
In the particular case of a binary classification, the equation simplifies as:

\begin{equation}
H_p = - \frac{1}{N}\sum_{i=1}^{N}y_i\log{(p_i)} + (1-y_i)\log{(1-p_i)}.
\end{equation}

\subsection{Bayesian Information Criteria}

 Bayesian information criteria (BIC)  can be used as a  technique that penalizes the likelihood in model selection \cite{Schwarz1978}, where the lower the value of BIC, the better the result. The BIC is given  by 
\begin{equation}
BIC = - 2\log \mathcal{L} + k\log{n},
\end{equation}
where $\mathcal{L}$ is the maximum value of the likelihood,  $k$ is the number of free parameters in the  model, and $n$ is the number of observations.

\subsection{Akaike  Information Criteria}

Akaike  Information Criteria (AIC), named after Hirotugu Akaike is given  by 
\begin{equation}
AIC = - 2\log \mathcal{L} + 2k,
\end{equation}
where $\mathcal{L}$ is the maximum value of the likelihood,  and  $k$ is the number of estimated parameters. Similar to  BIC, the preferred model is the one with the minimum AIC value.

\section{Results}

This section describes the relationships between  pregnancy occurrence and  outcomes against women's age and tubal anatomic properties. For that we utilise partial dependence plots (PDP) \cite{friedman2001,Brandon2017}. PDPs are useful to visualise the relationship between a subset of the features  and the response while accounting for the average effect of the other predictors in the model. 

\subsection{Model fit}

\begin{figure*}
\includegraphics[width=0.325\linewidth]{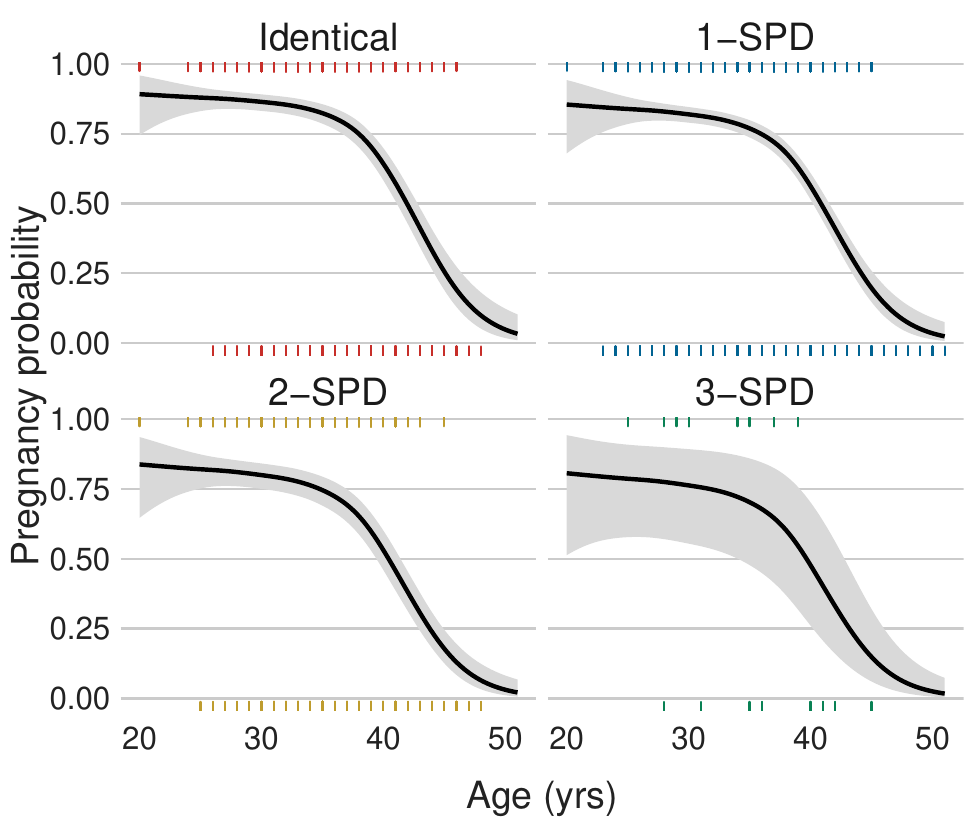}
\includegraphics[width=0.325\linewidth]{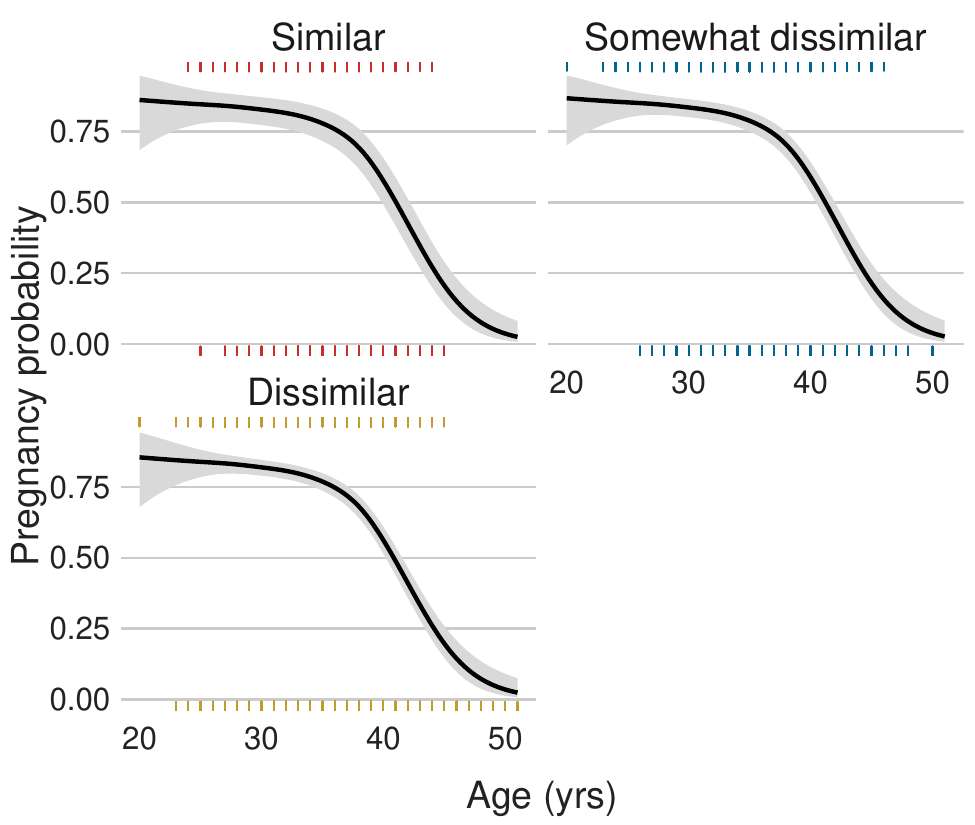}
\includegraphics[width=0.325\linewidth]{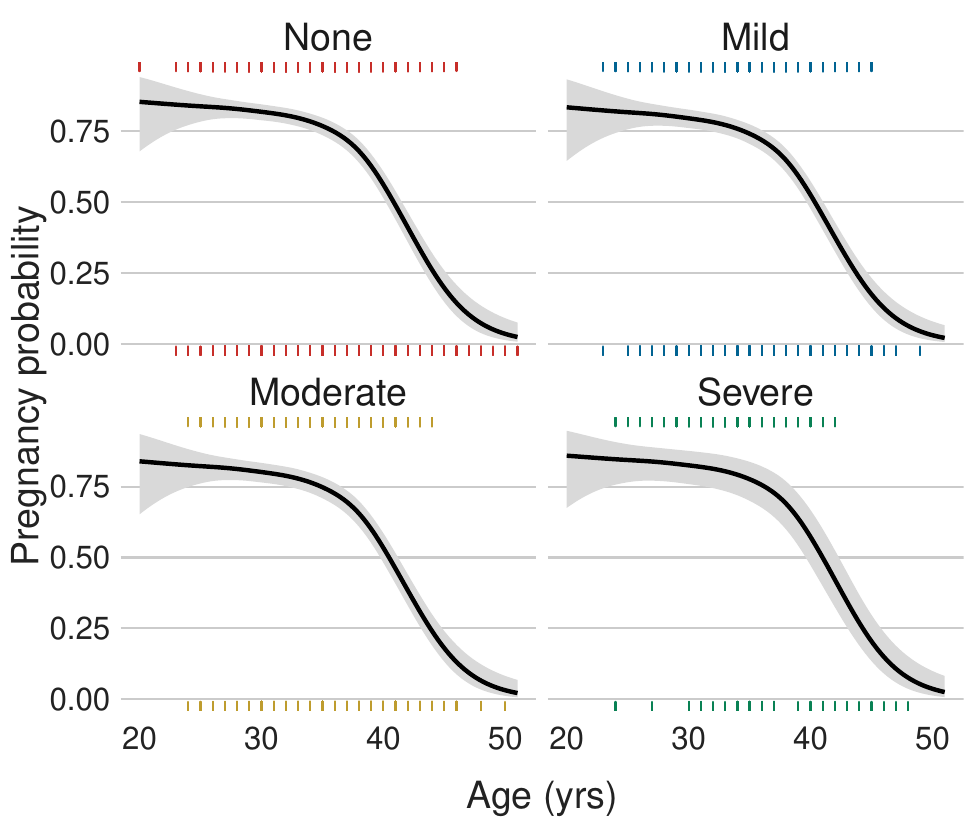}
\caption{Likelihood of pregnancy by age for each of the observed anatomic properties. Location, left panel; Diameter, middle panel; Fibrosis, right panel. The shaded grey areas depict 95\% confidence intervals. }
\label{fig:age_mc}
\end{figure*}

\begin{figure*} 
\includegraphics[width=0.325\linewidth]{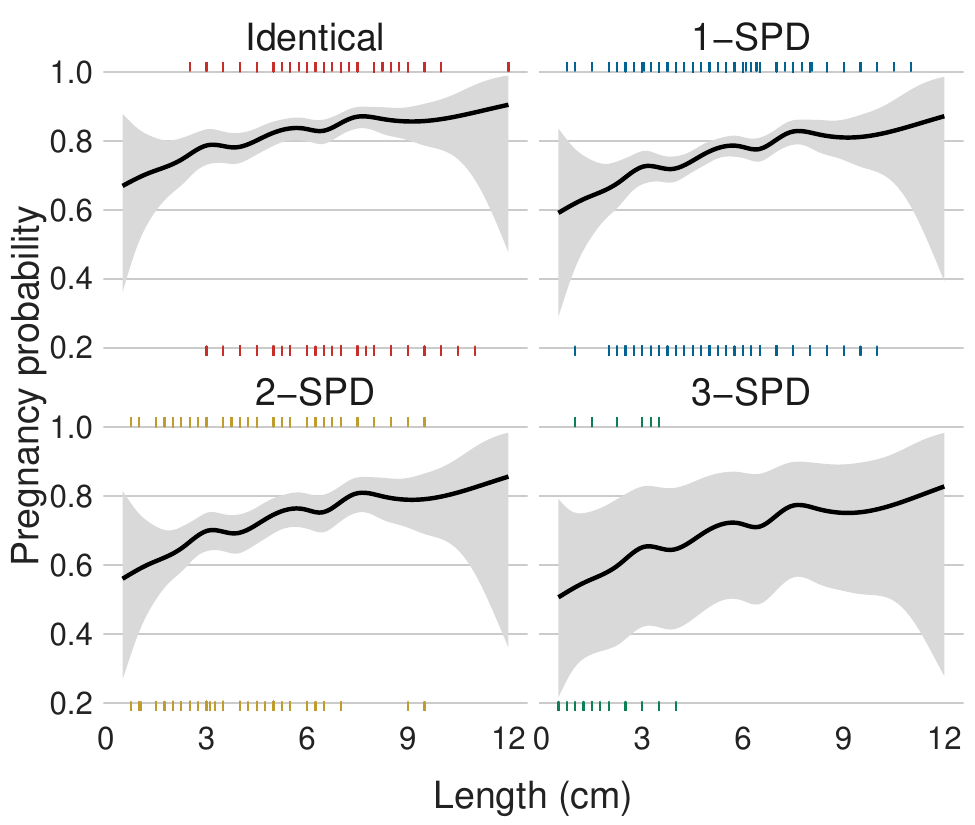}
\includegraphics[width=0.325\linewidth]{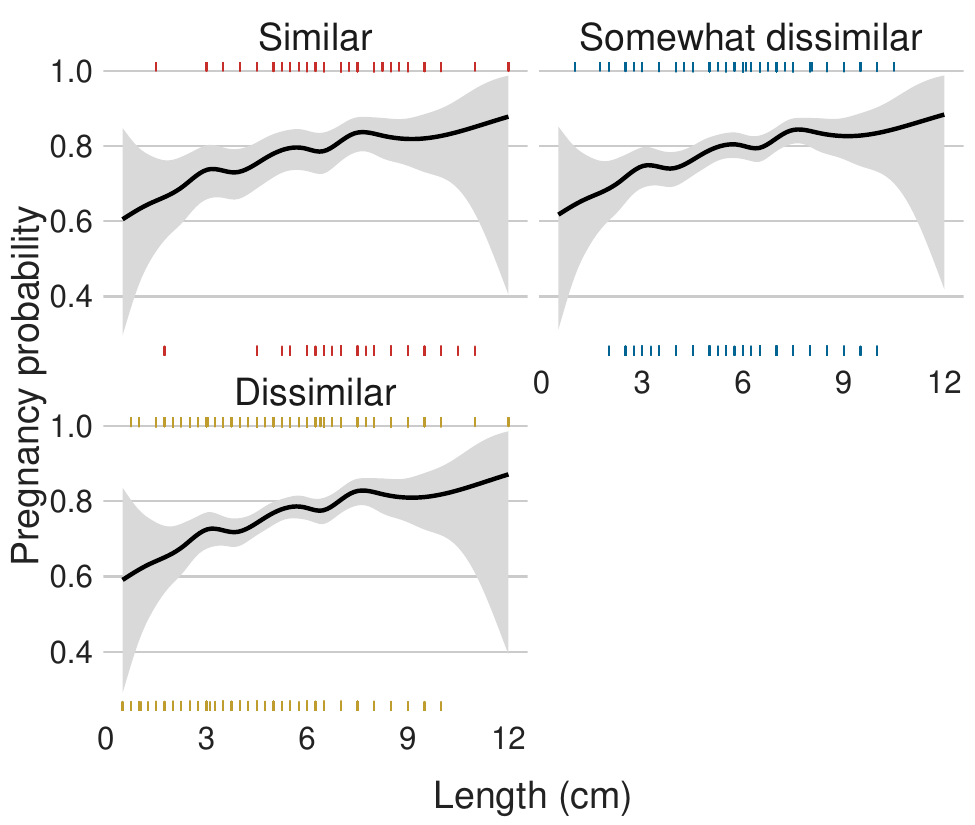}
\includegraphics[width=0.325\linewidth]{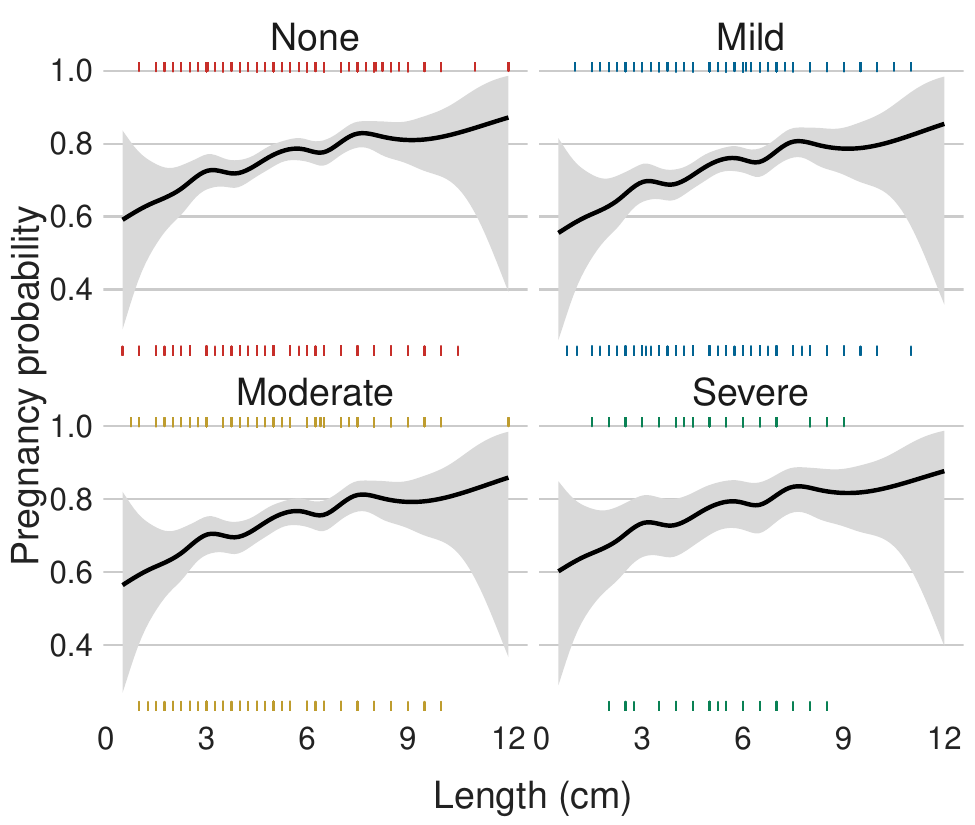}
\caption{Likelihood of pregnancy by tubal length for each of the observed anatomic properties. Location, left panel; Diameter, middle panel; Fibrosis, right panel. The shaded grey areas depict 95\% confidence intervals. }
\label{fig:tl_mc}
\end{figure*}

Figs. \ref{fig:age_mc} and \ref{fig:tl_mc} display the conditional fits for pregnancy likelihood in terms of women's age and tubal length for each of the categorical anatomic properties. The shape and steepness of the curves  are indicators of the predictor's relative influence.

Fig. \ref{fig:age_mc} shows the predominant role of age in pregnancy occurrence, which is consistent regardless of women's different anatomic properties. Pregnancy probability declines steadily as age increases, the rate of decline increasing sharply after $\approx 35$ years of age. The higher uncertainty for women with 3-SPD location is due to the small number  in this group, as can be seen in Fig. \ref{fig:ana_mosaic}. 

Tubal length is another good predictor of pregnancy likelihood, as portrayed in Fig. \ref{fig:tl_mc}. The longer the fallopian tube, the higher the pregnancy odds. This trend is also independent of other properties.  

To summarize, age and tubal length are the major factors determining the odds of pregnancy.

\begin{figure*} 
\includegraphics[width=0.45\linewidth]{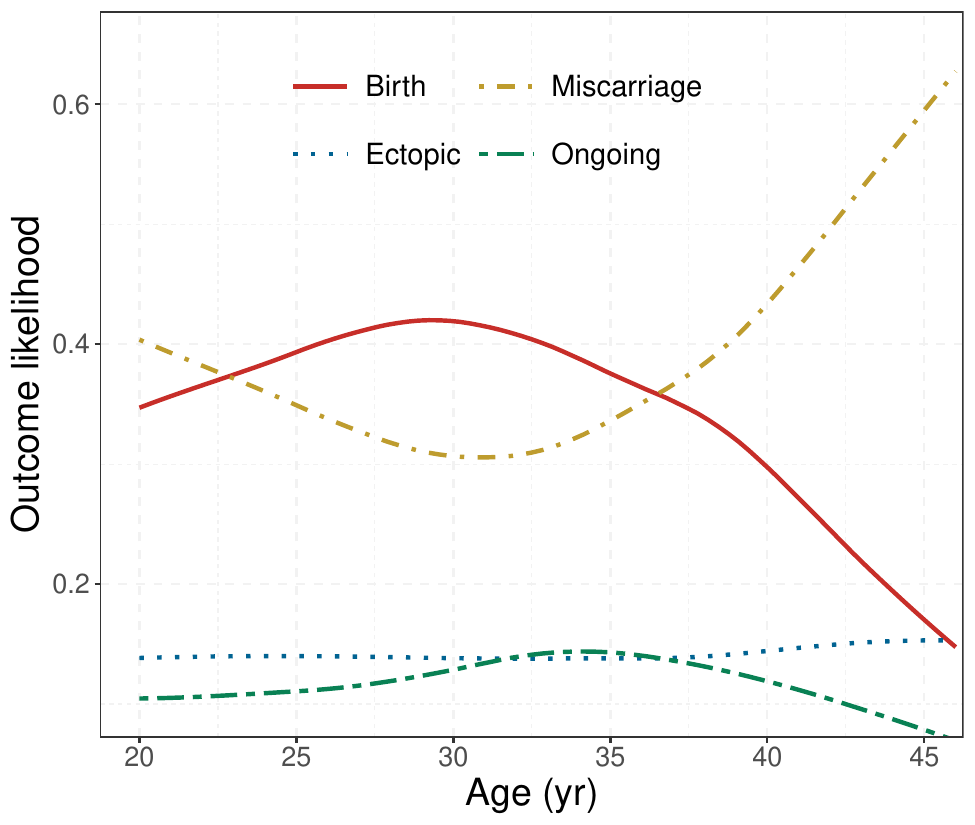}
\includegraphics[width=0.45\linewidth]{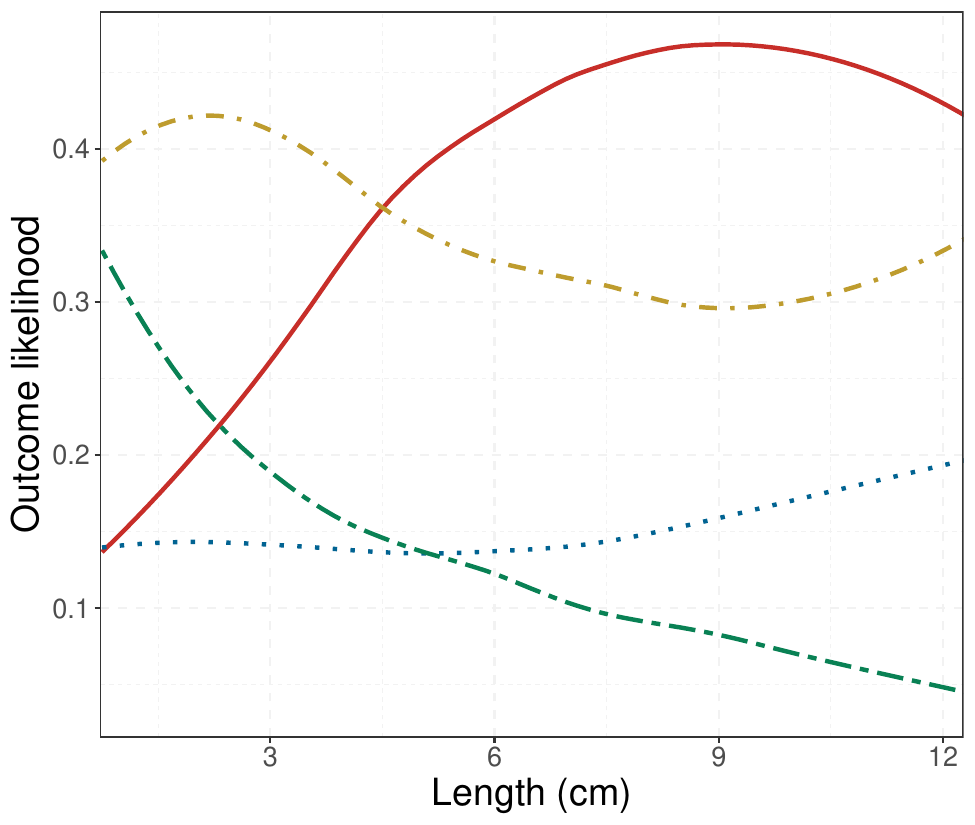}
\caption{Likelihood of pregnancy outcomes (birth, miscarriage, ectopic, ongoing) by age in the left panel and by tubal length in the right panel from the  multinomial GAM model. }
\label{fig:pdpout}
\end{figure*}

Fig. \ref{fig:pdpout} shows the trends between outcomes likelihood as a function of age (left plot) and tubal length (right plot).  The left plot quantifies the well known association between age and birth odds. Birth likelihood peaks by age 30 with the chances dropping sharply in the 40s as the likelihood of miscarriage rises with advancing age. The ectopic and ongoing pregnancy groups show no clear relationship with age. The right panel shows the relationship of tubal length to pregnancy outcome. As tubal length increases, the probability of birth rises dramatically; conversely, the odds of both miscarriage and ectopic pregnancy decline.

From Fig.  \ref{fig:pdpout}, we can conclude that women at ages $\lesssim$ 30 yrs and tubal length $\gtrsim$ 5 cm have the highest odds of a successful pregnancy. On the other hand, women $\gtrsim$ 40 yrs old and tubal lengths ( $\lesssim$ 5 cm) have the highest risk of miscarriage or  ectopic pregnancy.
%,  with their chances of giving birth dropping  by 2/3 or more.  

\subsection{Feature relevance}

\begin{figure} 
\includegraphics[width=0.95\linewidth]{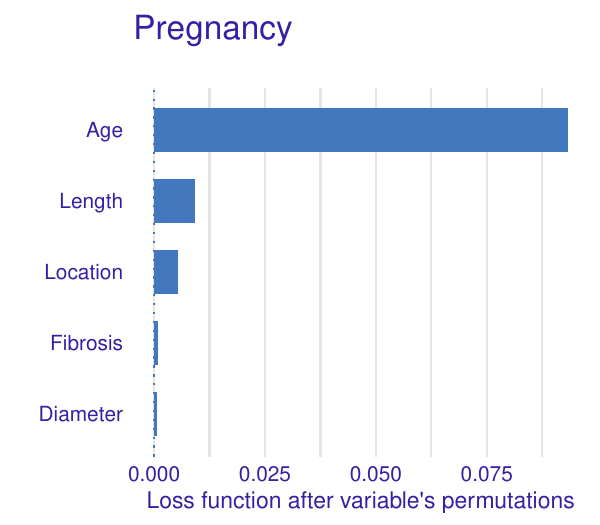}
\caption{Age and tubal anatomic properties ranked  according to their influence on pregnancy likelihood. The x-axis depicts the loss-drop based on the cross-entropy loss function for each feature.}
\label{fig:virlik}
\end{figure}

\begin{figure*}
\includegraphics[width=0.245\linewidth]{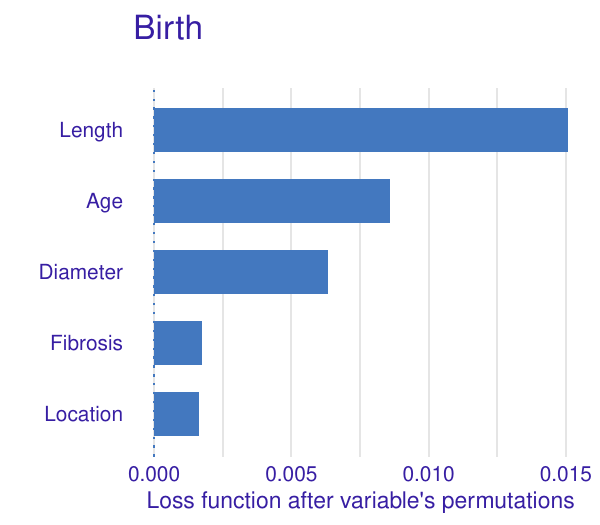}
\includegraphics[width=0.245\linewidth]{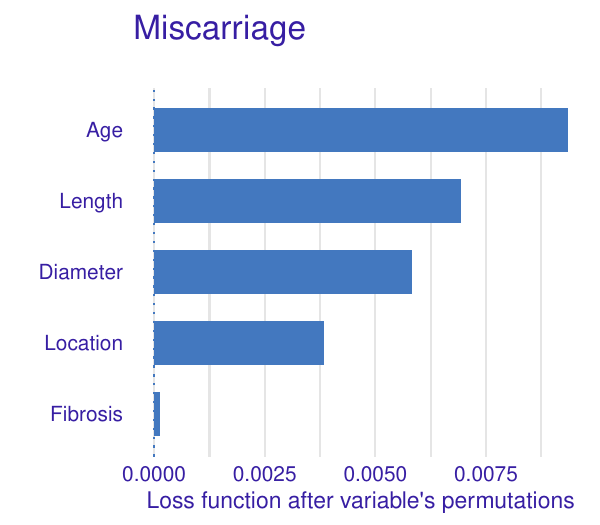}
\includegraphics[width=0.245\linewidth]{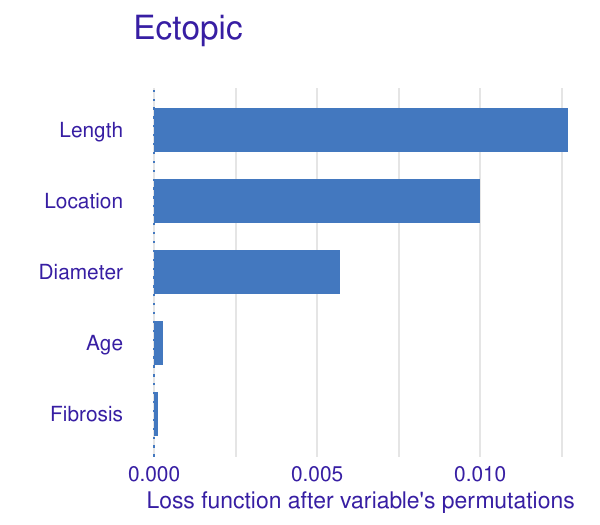}
\includegraphics[width=0.245\linewidth]{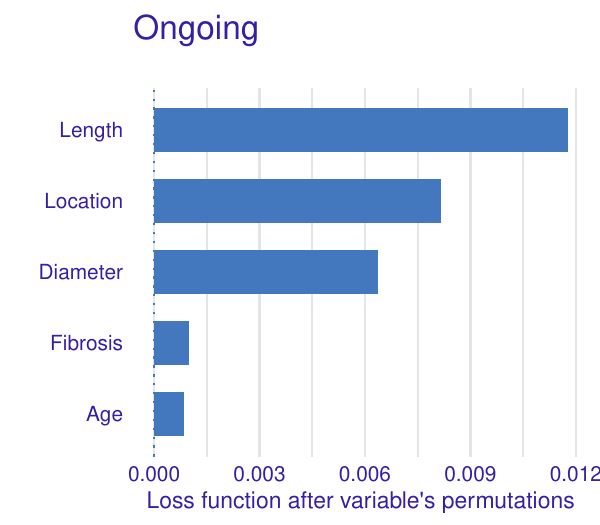}
\caption{Women's age and anatomic properties ranked  according to their influence on each pregnancy outcome. The x-axis depicts the loss-drop based on the cross-entropy loss function for each feature.}
\label{fig:virout}
\end{figure*}

A more formal approach to evaluate the importance of the women's multivariate and interrelated characteristics  is to use the cross-entropy loss  for assessing the significance of each property  in the model after taking the others into account.  Specifically, the idea is to  quantify the  hypothesis that a given woman's feature  has no influence on the probabilistic threshold above which pregnancy or a given pregnancy outcome can occur.  The influential rank of each property for pregnancy likelihood  is shown in Fig. \ref{fig:virlik} and for pregnancy outcome likelihood in Fig. \ref{fig:virout}.

Fig. \ref{fig:virlik} shows that in predicting pregnancy, age and  tubal length are by far the most influential properties, followed by anastomosis location. Diameter differences and fibrosis play less important roles.

Fig. \ref{fig:virout} portrays the influential properties for each of the pregnancy outcome categories. Tubal length and age are the most influential  factors related to the odds of birth and of miscarriage which are correlated inversely.  Tubal length is also the most influential factor for predicting a possible ectopic pregnancy. The shorter the fallopian tube, the higher the odds of this event.  The dominant influence of tubal  length  in the ongoing pregnancy group requires caution. In these cases the actual outcome is unknown and there may be an unknown bias in this class.

\subsection{Model evaluation}

This section compares the predictive power and model performance based on anatomic properties against a benchmark model based on sterilization method. The rationale behind this is to determine if knowing the postoperative anatomic properties improves the model's capability to predict the likelihood of pregnancy and subsequent outcomes. 

The following 3 scenarios were tested: A) age and sterilization method;  B) age and tubal anatomy; C) age, sterilization method, and tubal anatomy.

%The performance of each model was evaluated using cross-entropy loss ($H_p$). 
The results are displayed in table \ref{tab:metric}.

\begin{table}[ht]
\caption{Bayesian information criteria (BIC), and  Akaike information criteria (AIC) for the following scenarios A) age and sterilization method;  B) age and tubal anatomy; C) age, sterilization method, and tubal anatomy. The middle column refers to pregnancy likelihood (PL) and the right-hand column to outcome likelihood (OL).\\}
\label{tab:metric}
\centering
\begin{tabular}{|c|l|l|l|l| }
 \hline
&  \multicolumn{2}{c|}{\textbf{BIC}} & \multicolumn{2}{c|}{\textbf{AIC}} \\
 \cline{2-5}
Scenario & PL &  OL & PL &  OL \\
 \hline
  A & 6363   & 17390 & 6312 & 17129 \\
  B & 6329  & 17288 & 6238 & 16951 \\
  C & 6350 & 17436 & 6240 &  16936\\
 \hline
\end{tabular}
\end{table}

For predicting both pregnancy likelihood and outcome likelihood, Model B (age and tubal anatomy) outperforms Model A (age and sterilization method) and Model C (age, sterilization method, and tubal anatomy).  While in the case of predicting  pregnancy outcomes, there is a slightly  preference for the model in which the sterilization method information is also provided, according to AIC.

\section{Comment}

Previous studies have shown age at surgery to be the primary factor associated with the success of female sterilization reversal. Some studies have examined postoperative tubal length and a few have examined the effect of anastomosis location on pregnancy and birth rates; the results have been conflicting \cite{van2017}.

Previous studies have had insufficient patient numbers to analyze multiple factors simultaneously. Our study, with its large patient population, permits more in-depth analysis and examines  aspects of postoperative tubal anatomy not reported before. 

Preoperative counseling and obtaining informed consent for tubal anastomosis requires discussing the chances of success and of risks, including possible ectopic pregnancy. To assist with this, a previous report from the Tubal Surgery Database  \cite{BERGER2016} presented tables of pregnancy and pregnancy outcome rates stratified by age and sterilization method. Both of these factors are known before reparative tubal surgery. 

After tubal reversal, patients usually ask how their findings affect their chances of having a baby. Reframing the question, does tubal anatomic information add predictive power for pregnancy and pregnancy outcome probabilities compared with what was known before surgery? That is the question this study was undertaken to answer.

Tubal anatomy observed at reversal surgery is related to the method of sterilization. Clips are the least damaging, while coagulation can be extensively damaging to the tubes. Tubal anatomic parameters (length, anastomosis location, segment diameters, fibrosis) are interrelated. Anatomy often differs between right and left tubes in individual patients. Furthermore, it is unknown which of the tubes is involved in a given pregnancy. These issues have not been addressed in previous studies, yet they are important to reflect the realities of clinical medicine. The statistical methodology we have described addresses these issues.

Of the models we tested, the one with anatomic properties gives a better fit to the data than the one without. We believe this model to be robust based on the consistency of trends in the exploratory analysis, visualization of the fit, and variable importance analysis.

Limitations of this study include the absence of data about other factors that contribute to fertility, such as semen analysis; frequency and timing of intercourse; and ovulation history.  Nevertheless, our study probes deeper and with greater statistical power than past studies of fertility after tubal reversal surgery.

\section{Conclusions}

 Information  about tubal anatomy acquired at reversal surgery adds predictive
power for pregnancy and pregnancy outcome probabilities compared with what was known before surgery. This is especially helpful during postoperative counseling when addressing patients' questions about prognosis.

\section*{Conflict of interest}
All authors approved the final manuscript as submitted and agreed to be accountable for all aspects of the work. The corresponding author attests that all listed authors meet authorship criteria and that no others meeting the criteria have been omitted.

%

%%Harvard (name/date)
\bibliographystyle{SageV}
%%Vancouver (numbered)
%\bibliographystyle{SageV}
\bibliography{sample.bib}

\end{document}